\documentclass[cameraready]{Interspeech}
\usepackage{graphicx} % Required for inserting images
\usepackage{caption}
\usepackage{cite}

\title{Membership Inference Attacks against Large Audio Language Models}
\author[affiliation={1}]{Jia-Kai}{Dong}
\author[affiliation={1}]{Yu-Xiang}{Lin}
\author[affiliation={1,2}]{Hung-yi}{Lee}

\address{
    $^1$ National Taiwan University \\
    $^2$ NTU Artificial Intelligence Center of Research Excellence 
}

\email{b11901067@ntu.edu.tw}

\usepackage{comment}

\begin{document}

\maketitle

\begin{abstract}
We present the first systematic \textbf{membership inference attack (MIA)} evaluation of LALMs. Using \textbf{Multi-modal Blind Baselines} based on textual, spectral and prosodic features, we demonstrate that common audio datasets exhibit near-perfect train/test separability (AUC $\approx 1.0$) even without model inference, thus MIA may primarily detect distribution shift. 
We therefore introduce a blind-baseline protocol to control for this confound.
Under this protocol, we identify that the distribution-matched datasets enable reliable MIA evaluation without distribution-shift artifacts. 
We benchmark multiple MIA methods and conduct modality disentanglement experiments on these datasets.
The results reveal that LALM memorization is \textbf{cross-modal}, arising only from binding a speaker’s vocal identity with its text. These findings establish a principled standard for auditing LALMs beyond spurious correlations. Our codebase is available at \url{https://github.com/snooow1029/ALM_MIA}.

\end{abstract}

% The privacy implications of Large Audio Language Models (LALMs) remain largely unexplored. Unlike text, audio data encodes abundant non-semantic information, including recording conditions and speaker prosody, which introduces inherent distribution shifts between training and test sets. In this work, we present the first systematic Membership Inference Attack (MIA) evaluation of LALMs under a realistic \textbf{two-stage generation protocol}. We identify a validity crisis in current audio privacy assessments: standard MIA metrics predominantly detect acoustic artifacts rather than true model memorization. By extending a rigorous \textbf{multi-modal blind baseline} framework that incorporates spectral and prosodic representations, we show that widely used speech benchmarks exhibit near-perfect train/test separability, with AUC values approaching 1.0 even without any model inference. Moreover, MIA scores on these datasets are strongly correlated with blind acoustic features ($r > 0.7$), rendering them unreliable as privacy indicators. To establish a principled auditing methodology, we focus on distribution-matched datasets and employ \textbf{modality disentanglement}. Our results reveal that memorization in LALMs manifests as \textbf{cross-modal binding}, requiring the specific pairing of a speaker’s vocal identity with its corresponding textual content. These findings set a new standard for auditing LALMs by moving beyond spurious correlations and isolating the true nature of multimodal memorization.

\section{Introduction}
Large Audio Language Models (LALMs) ~\cite{ghosh2025audio,liu2025seallmsaudiolargeaudiolanguagemodels,rubenstein2023audiopalmlargelanguagemodel,lu2026desta2,ghosh2026audio,chu2023qwenaudioadvancinguniversalaudio,tang2024salmonn} have recently emerged as a powerful paradigm for multimodal reasoning, unifying tasks such as speech translation, sound event detection, and music understanding within a single framework. While these models achieve impressive generalization by training on massive web-scale datasets, they raise critical concerns regarding privacy and copyright, particularly whether sensitive audio content is memorized during training. 

\textbf{Membership Inference Attacks (MIA)}, which aim to determine whether a specific sample was used during model training, have become a standard tool for auditing such risks \cite{carlini2021extracting,maini2024llmdatasetinferencedid}. 
Although MIA has been extensively studied in text-only LLMs~\cite{mattern2023membershipinferenceattackslanguage,duan2024membershipinferenceattackswork, fu2024membership, wen2024membership, galli2024noisy, song2025mias, zawalski2026detecting} and VLMs~\cite{li2024membershipinferenceattackslarge,hu2025membershipinferenceattacksvisionlanguage}, its applicability to LALMs remains unexplored. Existing privacy studies in the audio domain focus mainly on representation learning or speaker recognition systems ~\cite{tseng2022membership, Chen_2024}, where attacks target fixed-dimensional embeddings rather than sequence-level generation. 

In contrast, LALMs introduce a distinct privacy setting because they are trained on large-scale audio–text corpora that couple acoustic characteristics with linguistic content. Although limited data exposure (often a single epoch) is commonly believed to reduce verbatim memorization~\cite{duan2024membershipinferenceattackswork}, LALMs may still retain cross-modal audio–text associations during training. Consequently, membership leakage may stem from memorized audio–text associations. For example, such leakage may reveal whether a particular speaker–content pair was present in the training data, posing privacy risks beyond conventional textual memorization.

% In contrast, LALMs are trained on massive web-scale corpora with limited exposure, often only a single epoch. 
% While this regime is commonly assumed to mitigate verbatim memorization~\cite{duan2024membershipinferenceattackswork}, the high capacity of LALMs still enables membership leakage through memorized \emph{cross-modal mappings} between acoustic signals and text. Such leakage reveals speaker–content associations, constituting a biometric privacy risk that is more granular and severe than textual duplication.
% Hence, membership leakage in LALMs may arise not only from memorized content, but also from memorized cross-modal associations between a speaker's voice characteristics and the corresponding textual content. Such leakage exposes speaker–content relationships, creating a biometric privacy risk that is potentially more sensitive than the textual memorization risks studied in prior LLMs.

% Auditing LALMs is further complicated by the continuous nature of audio, which encodes rich non-semantic cues such as speaker identity and recording conditions. 
However, detecting such memorization is challenging, as audio contains rich non-semantic cues, including speaker identity and recording conditions.
Consequently, privacy threats in LALMs are dominated not by copyrighted content, but by irreversible identity–content binding. Moreover, standard audio benchmarks often exhibit acoustic distribution shifts between training and test sets that can spuriously inflate MIA performance. As demonstrated in LLMs \cite{das2025blindbaselinesbeatmembership}, simple \emph{blind baselines} that exploit such dataset artifacts can outperform sophisticated MIAs, raising concerns about whether reported leakage truly reflects memorization  ~\cite{puerto2025scaling}.

In this paper, we conduct the first systematic MIA study on two state-of-the-art open-source models: Audio-Flamingo 3 (AF3) ~\cite{ghosh2026audio} and Music-Flamingo (MF)~\cite{ghosh2025musicflamingoscalingmusic}. 
% The transparent data provenance of these models enables a direct audit of the foundation weights themselves. 
% This approach sidesteps the use of computationally prohibitive and often unfaithful surrogate ``shadow'' models, enabling an authentic evaluation of memorization across eight diverse datasets spanning ASR, Audio Captioning, and Music Understanding. 
Because their training data are fully documented, ground-truth membership labels are available, eliminating the need for computationally expensive and often unfaithful surrogate “shadow” models. This enables a reliable evaluation of memorization across eight diverse datasets spanning ASR, audio captioning, and music understanding.

Furthermore, to disentangle genuine memorization from dataset-induced artifacts, we introduce a \textbf{Multi-modal Blind Baseline} framework that quantifies distribution shifts across metadata, textual content, and acoustic features.

Our contributions are threefold:
\begin{itemize}
    \item We present the first comprehensive benchmark of sample-level MIA for LALMs in an audio/text-to-text setting. We evaluate seven confidence-based MIA attacks methods across eight audio-centric tasks.
    \item We propose a rigorous \textbf{Multi-modal Blind Baseline} protocol to audit privacy risks arising from dataset design. We show that common benchmark practices introduce severe acoustic distribution shifts that inflate MIA AUCs and strongly correlate with blind acoustic classifiers.
    \item On distribution-matched datasets, we show that LALM memorization is \textbf{cross-modal}, depending on the specific pairing of acoustic attributes such as speaker identity and text. Privacy risks are thus highest when sensitive content is tightly linked to a particular speaker.
\end{itemize}

Overall, our findings highlight the fragility of current privacy assessments in the audio domain and underscore the importance of controlling for acoustic distribution shifts when auditing LALMs. We argue that reported MIA results should be interpreted in conjunction with blind baseline diagnostics; without such analysis, it is difficult to disentangle genuine memorization from distributional artifacts.

\section{Methodology}
\label{sec:method}

\begin{figure}[t]  
  \centering
  \includegraphics[width=\linewidth]{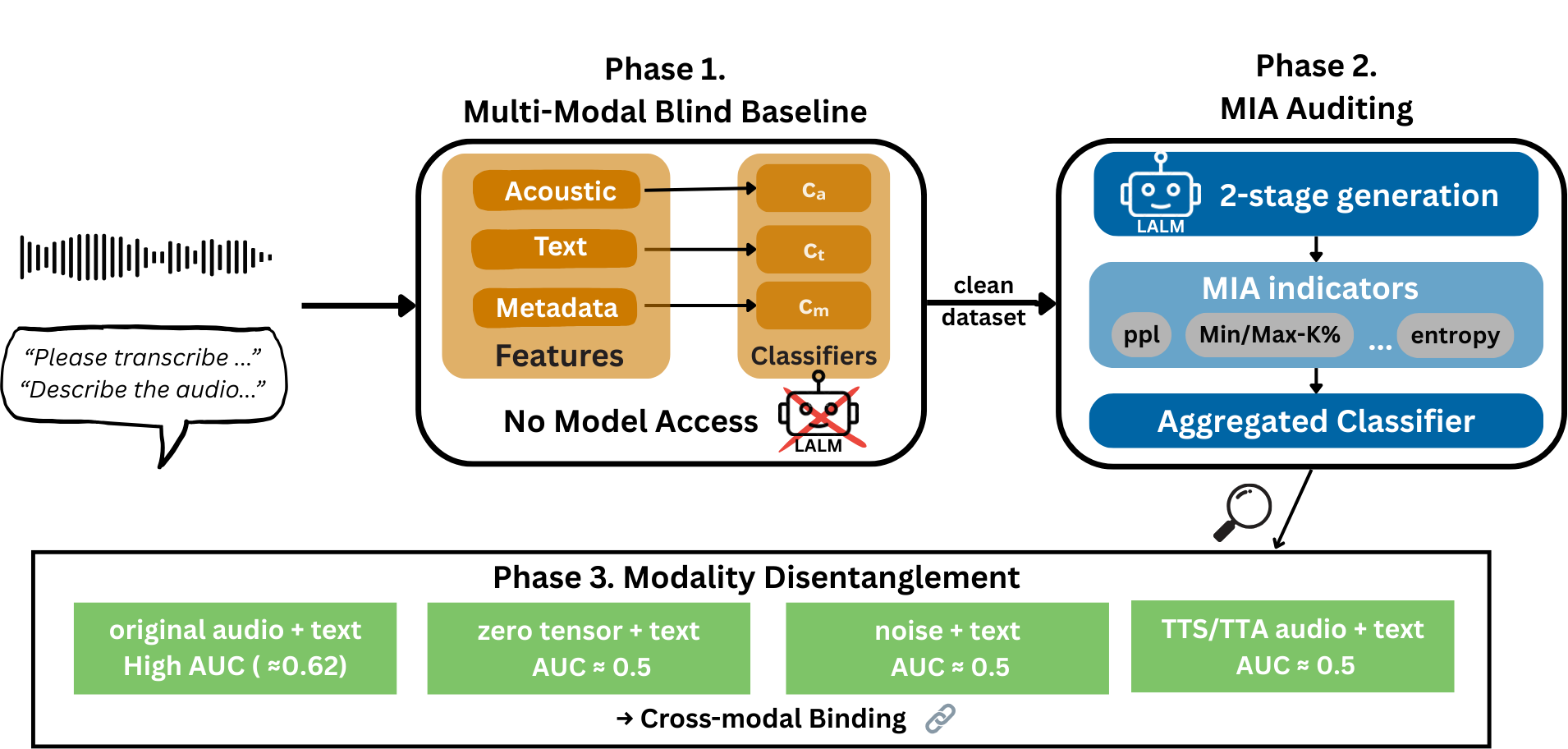}
  \caption{Overview of the 3-phase LALM privacy auditing framework: 
  \textbf{(1) Multi-Modal Blind Baseline:} Filters out samples with confounding distributional shifts to yield a clean dataset. 
  \textbf{(2) MIA Auditing:} Audits the target LALM on the clean dataset using membership indicators from a two-stage generation process. 
  \textbf{(3) Modality Disentanglement:} Probes confirmed memorization to identify and characterize cross-modal binding.}
  \label{fig:framework}
\end{figure}

We propose a multi-phase privacy auditing framework, as illustrated in Figure~\ref{fig:framework}. 
Our approach begins by establishing a \textbf{Multi-modal Blind Baseline} (Phase 1) to quantify any distribution shifts inherent in the data, without accessing the target LALM. 
We then perform the primary \textbf{MIA Auditing} on the model (Phase 2) and, crucially, correlate its success with the blind baseline's performance. 
This diagnostic step allows us to disentangle spurious MIA success driven by data artifacts from genuine model memorization. 
Finally, for cases of validated memorization, we conduct \textbf{Modality Disentanglement} experiments (Phase 3) to characterize the memory's nature, specifically testing for the existence of \textbf{cross-modal binding}.

These experiments are designed to test the hypothesis that privacy leakage in LALMs is driven by \emph{cross-modal binding}, where memorization is only triggered when the model encounters the precise pairing of original acoustic features and their corresponding textual sequences.

\subsection{Inference Protocol: Two-Stage Generation}
To simulate a realistic gray-box auditing scenario where precise ground-truth transcripts used during training may be unavailable, we adopt a \textbf{Two-Stage Generation} protocol:

\noindent\textbf{Stage 1: Autonomous Decoding.}\quad
Given an audio input $x$, the target LALM performs greedy decoding to produce a self-generated output sequence $\hat{y}$.

\noindent\textbf{Stage 2: Self-Conditioned Scoring.}\quad
We treat $\hat{y}$ as pseudo-ground-truth and re-run the forward pass to obtain token-level logits corresponding to $\hat{y}$.
Membership metrics are then computed based on these self-conditioned probabilities.

\subsection{Phase 1: Multi-modal Bias Audit Protocol}\label{sec:feature_sets}To prevent spurious MIA success and ensure the validity of privacy assessments, we first propose a systematic \textbf{Multi-modal Blind Baseline} framework. This protocol identifies distributional shortcuts that models may exploit across two hierarchical levels: (1) \textbf{Inter-dataset Shift}, where mismatches between disparate data sources cause MIA metrics to act as trivial domain classifiers; and (2) \textbf{Intra-dataset Shift}, where subtle statistical discrepancies arise between the training and test splits of the \textit{same} dataset. In the audio domain, such shifts are particularly common, as train and test splits are often constructed using different curation or preprocessing pipelines; for example, in \textbf{GigaSpeech}~\cite{chen21o_interspeech}, transcription standards lead to systematic differences between splits.
Our protocol quantifies these shifts by training \emph{blind} Logistic Regression classifiers on data-intrinsic features without accessing the LALMs. To isolate the contribution of different sources of distribution shift, we construct three blind baselines at increasing levels of modality coverage:

\begin{itemize}
    \item \textbf{Metadata}: Structural statistics including audio duration, file size, and word/character counts.
    \item \textbf{Text}: Utterance-level TF-IDF representations (unigrams and bigrams).
    \item \textbf{Acoustic}: Fixed-dimensional utterance-level descriptors obtained by aggregating frame-wise low-level acoustic features, including MFCCs, spectral statistics (centroid, bandwidth, rolloff), pitch, energy (RMS), and zero-crossing rate. These features are specifically designed to capture recording conditions, channel characteristics, and dataset-specific acoustic artifacts rather than linguistic content.
\end{itemize}

All features are aggregated into fixed-dimensional, sample-level representations. We adopt simple Logistic Regression classifiers to ensure that strong blind performance reflects salient distributional artifacts rather than classifier capacity.

We utilize the Pearson ($r$) correlation coefficient between these blind scores and the subsequent model-based MIA indicators (defined in Section~\ref{subsec:mia_metric}) as a diagnostic reference.\footnote{We also compute Spearman rank correlations as a robustness check and observe highly consistent diagnostic conclusions.} Rather than applying a fixed threshold, we argue that a high Blind Baseline AUC, when accompanied by a non-negligible positive correlation with the model's MIA scores, provides strong empirical evidence that the detected membership is likely confounded by distributional shortcuts. In such cases, the reported MIA performance may reflect the LALM's sensitivity to data-intrinsic artifacts rather than genuine memorization, thereby necessitating a more cautious interpretation of the privacy risks.

\subsection{Membership Inference Indicators}
\label{subsec:mia_metric}
We compute several complementary metrics from the model's output probabilities to serve as membership indicators:
\begin{itemize}
    \item \textbf{Perplexity (PPL)}: The average negative log-likelihood (NLL) of the sequence $\hat{y}$ ~\cite{8429311}.
    \item \textbf{Shannon Entropy}: The average entropy of the predicted token distributions, measuring predictive certainty ~\cite{6773024}.
    \item \textbf{Min-k\% / Max-k\% Prob}: Average NLL of the $k\%$ tokens with \textit{lowest} and \textit{highest} probabilities, respectively~\cite{shi2024detectingpretrainingdatalarge, zhang2024min}.
    \item \textbf{Max-Rényi MIA}: Measures probability mass concentration via Rényi divergence with orders $\alpha \in \{0,1,2,\infty\}$~\cite{li2024membershipinferenceattackslarge}.
    \item \textbf{Zlib Ratio}: The NLL normalized by the zlib compression size of the text ~\cite{8429311}.
    \item \textbf{Max Probability Gap}: The sequence-averaged gap between top-1 and top-2 token probabilities, capturing over-confident predictions indicative of memorized tokens~\cite{li2024membershipinferenceattackslarge}.
\end{itemize}
Following \cite{maini2024llmdatasetinferencedid}, we concatenate these metrics into a feature vector and train a Logistic Regression classifier using stratified 10-fold cross-validation within each dataset to produce membership probabilities.

\subsection{Modality Disentanglement Protocol}
For datasets passing the bias audit (Blind AUC $\approx 0.5$), we probe the mechanistic nature of memorization via modality disentanglement. We ask whether the membership signal arises from \textbf{cross-modal binding}, namely the memorization of a specific dependency between an acoustic realization and its textual content, rather than a simple textual prior. We evaluate the model under four configurations:

\begin{itemize}
    \item \textbf{Original}: Full acoustic and textual input, representing the matched training instance.
    \item \textbf{Text-Only (Silence)}: Audio input is replaced by a zero-tensor, isolating the \textit{textual prior} leakage.
    \item \textbf{Noise-Only}: Audio is replaced by Gaussian noise to check the impact of non-semantic acoustic activation.
    \item \textbf{Acoustic Resynthesis}: To isolate the impact of instance-specific acoustic features such as unique speaker identity or specific recording environments, we replace the original audio with a version synthesized from the textual content. For speech-centric datasets, we employ \textbf{Text-to-Speech (TTS)}; for environmental sound datasets, we utilize \textbf{Text-to-Audio (TTA)} generation. 
\end{itemize}
By comparing these conditions, we can quantify the contribution of \textbf{instance-specific acoustic features} to the overall membership signal. If the MIA performance collapses upon resynthesis or silence, it confirms that the model's memory is anchored to the unique binding of a specific speaker's voice to their utterance.

\section{Experimental Setup}
\label{sec:setup}

\begin{table}[t]
\centering
\caption{MIA performance (AUC) vs. blind baselines across datasets. Values in parentheses report Pearson correlation ($r_{\text{AF3}} / r_{\text{MF}}$) between LALM MIA and blind indicators. High correlation suggests spurious success driven by dataset artifacts.}
\label{tab:main_result}
\resizebox{\columnwidth}{!}{%
\begin{tabular}{lcc ccc}
\toprule
\textbf{Dataset} & \multicolumn{2}{c}{\textbf{MIA AUC}} & \multicolumn{3}{c}{\textbf{Blind Baseline: AUC ($r_{AF3} / r_{MF}$)}} \\ 
\cmidrule(lr){2-3} \cmidrule(lr){4-6}
& \textbf{AF3} & \textbf{MF} & \textbf{Metadata} & \textbf{Text} & \textbf{Acoustic} \\ 
\midrule
\textbf{Gigaspeech} & 87.6 & 90.5 & 71.9 (.59/.55) & 75.6 (.41/.50) & 78.9 (.52/.55) \\
\textbf{SPGISpeech} & 51.9 & 50.7 & 51.6 (.35/.33) & 50.0 (-.01/.06) & 52.0 (0.01/0.02) \\
\textbf{Librispeech} & 93.8 & 70.9 & 100.0$^{\dagger}$ (.75/.35) & 61.1 (.16/.11) & 99.8 (.78/.34) \\
\textbf{Tedlium} & 70.0 & 68.9 & 64.9 (.48/.53) & 85.7 (.24/.41) & 98.7 (.34/.30) \\
\textbf{VoxPopuli} & 62.0 & 68.1 & 55.1 (.24/.09) & 60.3 (.10/.13) & 66.7 (.07/.08) \\
\textbf{Clotho} & 52.4 & 48.9 & 51.1 (.01/.09) & 47.1 (-.01/.00) & 48.2 (.01/-.01) \\
\textbf{CochlScene} & 57.3 & 60.4 & 52.3 (.37/.21) & 65.3 (.34/.42) & 59.3 (.19/0.14) \\
\textbf{Nsynth} & 62.3 & 62.1 & 52.7 (-.03/-.09) & 61.7 (0.08/-.09) & 62.7 (.21/0.22) \\
\bottomrule
\multicolumn{6}{l}{\footnotesize $^{\dagger}$ Near-perfect baseline AUC indicates trivial train/test separability.}
\end{tabular}%
}
\end{table}

\subsection{Target Models}
While numerous LALMs have been proposed~\cite{ghosh2025audio,liu2025seallmsaudiolargeaudiolanguagemodels,rubenstein2023audiopalmlargelanguagemodel,lu2026desta2,ghosh2026audio,chu2023qwenaudioadvancinguniversalaudio,tang2024salmonn}, the majority of these models do not offer public access to their pre-training data, making precise membership auditing infeasible. We therefore evaluate our auditing protocol on two models with fully open-access resources: \textbf{Audio-Flamingo 3 (AF3)} \cite{ghosh2026audio} and \textbf{Music Flamingo (MF)} \cite{ghosh2025musicflamingoscalingmusic}, whose fully open-source weights and data provenance enable direct auditing at the foundation stage. This approach sidesteps the need for computationally expensive and often unfaithful surrogate shadow models, allowing an authentic evaluation of memorization across both general audio (AF3) and complex music (MF) domains.

\subsection{Dataset Selection}
We evaluate our auditing framework on eight datasets spanning three representative audio tasks, selected to cover diverse acoustic conditions, semantic structures, and data curation pipelines.

\begin{itemize}[noitemsep, topsep=0pt, leftmargin=*]
    \item \textbf{ASR}: \textbf{LibriSpeech}, \textbf{GigaSpeech}, \textbf{TED-LIUM}, \textbf{VoxPopuli}, and \textbf{SPGISpeech}, ranging from clean read speech to large-scale, noisy, and multi-speaker recordings, enabling analysis of both intra- and inter-dataset distribution shifts~\cite{7178964, chen21o_interspeech, hernandez2018ted, wang-etal-2021-voxpopuli, oneill21_interspeech}.
    \item \textbf{Audio Captioning}: \textbf{Clotho} and \textbf{CochlScene}, which emphasize non-linguistic acoustic semantics such as sound events and scenes, complementing speech-centric benchmarks~\cite{drossos2020clotho, jeong2022cochlscene}.
    \item \textbf{Music Synthesis}: \textbf{NSynth}, a musical instrument dataset used to analyze memorization in non-speech audio~\cite{engel2017neural}.
\end{itemize}

Members and non-members are defined by official training and test splits, respectively. For each dataset, we randomly sample a balanced cohort of 2000 to 5000 instances (comprising 1000–2500 pairs) to ensure statistical rigor and a consistent scale for our sample-level analysis. 

\subsection{Implementation Details} 

\noindent\textbf{Blind Baseline}\quad
The classifiers utilize the three feature sets described in Section~\ref{sec:feature_sets}, including utterance-level acoustic representations, TF-IDF textual features, and metadata statistics.

\noindent\textbf{MIA Indicator}\quad
For membership inference, as discussed in Section~\ref{subsec:mia_metric}, we aggregate various membership inference indicators into a 30-dimensional feature vector, where each dimension represents a distinct metric. 
Following \cite{maini2024llmdatasetinferencedid}, we sweep $k \in \{0.05, 0.1, \dots, 0.6\}$ for Min-k\% to capture low-probability tokens, and use $\alpha \in \{0, 1, 2, \infty\}$ for Max-Rényi to quantify probability concentration. Since individual indicators show inconsistent performance across diverse audio tasks, we adopt an aggregated design~\cite{maini2024llmdatasetinferencedid} to leverage complementary signals, as no single metric serves as a universal membership indicator.

\noindent\textbf{Training Protocol}\quad
Both the aggregated MIA and blind baseline classifiers are trained using stratified Logistic Regression. 
A linear Logistic Regression classifier is used to ensure detection reflects data artifacts rather than model capacity. We employ 10-fold cross-validation within each dataset to ensure every sample is evaluated as a test instance while strictly preventing cross-sample information leakage.

\noindent\textbf{Acoustic Resynthesis Models}\quad
To disentangle memorization effects from surface-level acoustic cues, we conduct an acoustic resynthesis ablation that replaces original audio signals with synthesized counterparts while preserving linguistic or semantic content.
For speech datasets such as VoxPopuli and SPGISpeech, we use \textbf{Cosyvoice2-0.5B}~\cite{du2024cosyvoice2scalablestreaming} with a standardized reference voice, where a small set of audio samples is selected from \textbf{LibriSpeech} as speaker reference audio to remove original speaker identities.
For audio captioning datasets including \textbf{Clotho}, we use \textbf{TangoFlux}~\cite{hung2024tangoflux} to generate synthetic soundscapes conditioned on the provided captions.
Due to the highly specialized nature of musical instrument synthesis and the brevity of labels, \textbf{NSynth} is excluded from the resynthesis ablation.
Three realizations per sample are generated per TTS/TTA setting; final results are reported as the average AUC to minimize stochastic variance.

\subsection{Evaluation Metrics}
To evaluate the effectiveness of the membership inference attacks, we treat the task as a sample-level binary classification problem. For each sample $i$ in a dataset $\mathcal{D} = \mathcal{D}_{mem} \cup \mathcal{D}_{non}$, we compute its membership score $S_i$ using the aggregated classifier or individual metrics. Membership performance is reported via the \textbf{Area Under the Receiver Operating Characteristic curve (AUC-ROC)}, which measures the probability that a randomly chosen member sample receives a higher score than a non-member. An AUC of 0.5 indicates random guessing, while 1.0 indicates perfect detection.

\section{Results and Discussion}

\begin{figure}[t]
    \centering
    \includegraphics[width=0.9\linewidth]{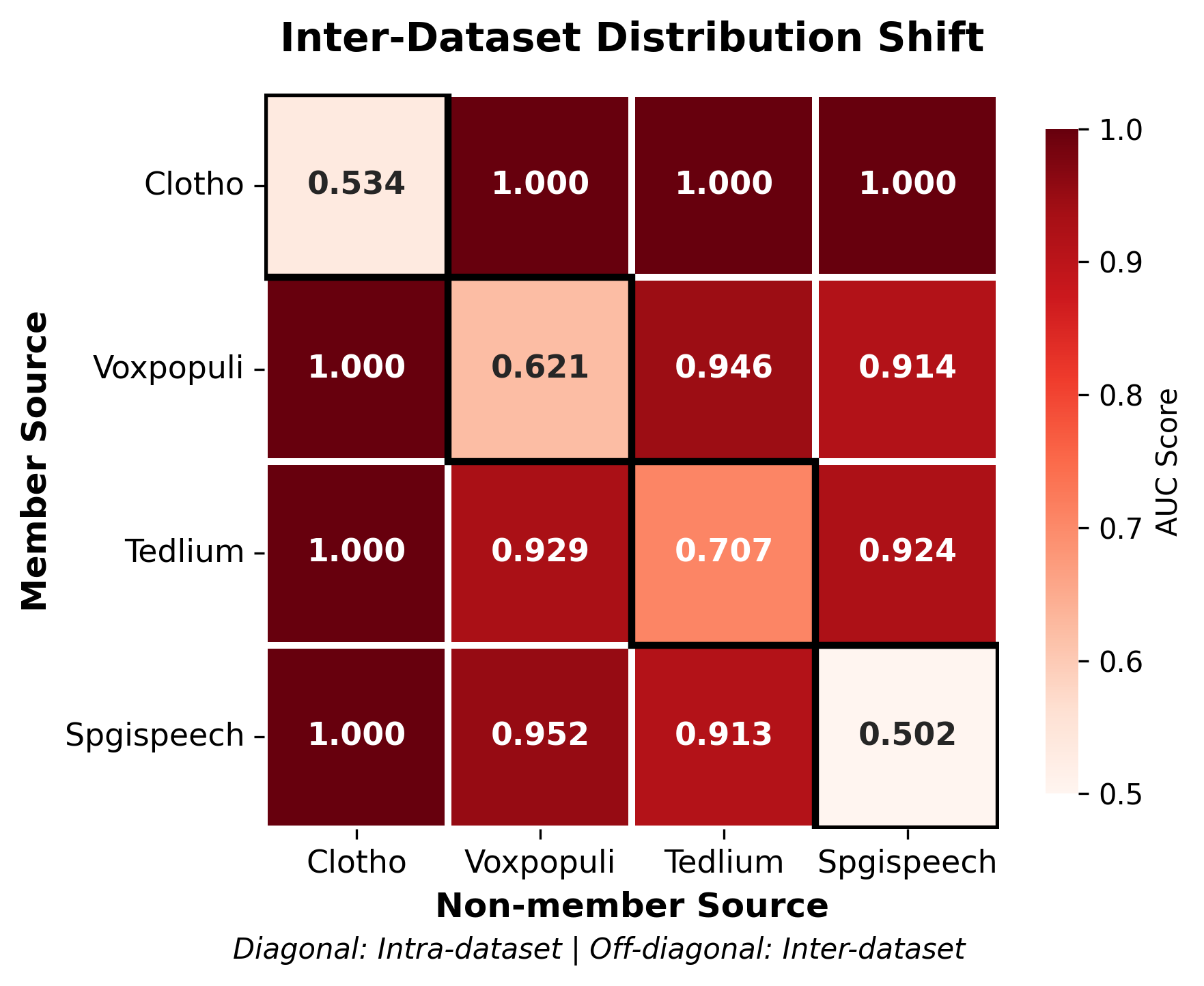}
    \caption{Cross-dataset MIA AUC heatmap. Diagonal: standard intra-dataset MIA (Train vs. Test). Off-diagonal: \textbf{membership-neutral pairings} (e.g., Dataset A Train vs. Dataset B Train), where near-perfect AUCs indicate domain shifts rather than genuine memorization.}
    \label{fig:inter_heatmap}
\end{figure}

\subsection{Spurious MIA Success and Distribution Shifts}
To ensure MIA reflects genuine memorization rather than data artifacts, we first isolate distributional shortcuts. Figure~\ref{fig:inter_heatmap} illustrates extreme \textbf{Inter-dataset Bias}: pairing member subsets across disparate sources (e.g., Clotho vs. SPGISpeech) yields AUCs of 1.0, proving MIA indicators often act as dataset classifiers rather than memorization detectors. 

Subtler \textbf{Intra-dataset Bias} also exists within common benchmarks. Table~\ref{tab:main_result} shows that datasets like \textbf{LibriSpeech} (AUC 93.8) exhibit high MIA scores that are largely deceptive. Our blind baseline reveals that LibriSpeech's acoustic features alone reach 99.8 AUC without model access. The strong correlation between MIA and blind scores ($r = 0.78$ for $M_1$). This confirms that much of the observed leakage in standard benchmarks is driven by systematic train-test discrepancies rather than genuine model memorization.

\subsection{Robustness on Distribution-Matched Datasets}
To establish a trustworthy audit, we focus on datasets where the Blind Baseline AUC $\approx 0.5$ (e.g., \textbf{SPGISpeech}, \textbf{Clotho}) in phase 1. We define these as ``clean'' datasets, as they ensure that observed MIA performance is not merely an artifact of distributional shifts. Under these controlled conditions, MIA performance collapses to near-random (AUC 50.7--52.4). This suggests that the evaluated LALMs exhibit strong sample-level privacy robustness under distribution-matched conditions. The lack of verbatim memorization indicates that the diverse acoustic realizations of a given text act as a natural regularizer, preventing the model from overfitting to specific sequences at the instance level. These findings indicate that current MIA methods perform poorly on datasets without distributional confounds.

\subsection{Cross-modal Binding}
While overall scores are low, modality disentanglement on the four datasets identified as relatively ``clean'' via our Phase 1 Blind Baseline protocol (\textbf{VoxPopuli}, \textbf{SPGISpeech}, \textbf{Clotho}, and \textbf{Nsynth}) reveals the mechanistic nature of LALM memory. As shown in Table~\ref{tab:disentanglement}, the membership signal in the Original condition collapses when modalities are decoupled.
Significance testing across multiple runs shows  statistically significant p-values ($p < 0.05$), confirming the empirical validity of the cross-modal binding effect. Replacing the original audio with \textbf{Silence}, \textbf{Noise}, or \textbf{TTS-Resynthesis} causes MIA AUCs to drop to near-random levels ($\approx 50\%$) across both models. These results suggest that LALMs do not memorize training data as isolated textual sequences or standalone acoustic fingerprints. Instead, memorization is \textbf{instance-specific and cross-modal}, emerging only from the \textbf{binding} between a specific vocal identity and its corresponding textual content.

\noindent\textbf{Privacy Implications} Our findings refine the threat model for audio privacy. The dominant risk is not the reproduction of verbatim text, but \textbf{linkage leakage}: the model's ability to associate a specific individual with a specific utterance. Breaking this binding via vocal anonymization or TTS offers a practical mitigation strategy for LALM training.

\begin{table}[t]
\centering
\caption{Modality Disentanglement (Sample-level AUC) on the four datasets identified as relatively ``clean'' via our Phase 1 Blind Baseline protocol. Results show a consistent collapse in MIA performance when acoustic or textual context is disrupted across both models. Subscripts in the Synthesized column denote standard deviation ($\pm$ SD) across multiple realizations.}
\label{tab:disentanglement}
\resizebox{\columnwidth}{!}{%
\begin{tabular}{l cc cc cc cc}
\toprule
\textbf{Dataset} & \multicolumn{2}{c}{\textbf{Original}} & \multicolumn{2}{c}{\textbf{Silence}} & \multicolumn{2}{c}{\textbf{Noise}} & \multicolumn{2}{c}{\textbf{Synthesized}} \\ 
\cmidrule(lr){2-3} \cmidrule(lr){4-5} \cmidrule(lr){6-7} \cmidrule(lr){8-9}
& $AF3$ & $MF$ & $AF3$ & $MF$ & $AF3$ & $MF$ & $AF3$ & $MF$ \\ 
\midrule
\textbf{VoxPopuli}  & \textbf{62.0}$^{\ast}$ & \textbf{68.1}$^{\ast}$ & 52.1 & 51.6 & 53.9 & 51.6 & $48.4_{\pm 2.1}$ & $49.8_{\pm 1.0}$ \\
\textbf{SPGISpeech} & \textbf{51.9}$^{\ast}$ & \textbf{50.7}$^{\ast}$ & 49.4 & 50.3 & 50.3 & 51.2 & $49.8_{\pm 0.2}$ & $49.4_{\pm 0.6}$ \\
\textbf{Clotho}     & \textbf{52.4}$^{\ast}$ & \textbf{48.9}          & 51.2 & 49.2 & 49.2 & 51.7 & $47.9_{\pm 0.6}$ & $50.0_{\pm 1.0}$ \\
\textbf{Nsynth}     & \textbf{62.3}$^{\ast}$ & \textbf{62.1}$^{\ast}$ & 50.0 & 48.1 & 48.8 & 49.0 & - & - \\
\bottomrule
\multicolumn{9}{l}{\footnotesize $^{\ast}$ Statistical significance ($p < 0.05$) comparing Original to all decoupled conditions under a paired $t$-test.}
\end{tabular}%
}
\end{table}

\section{Conclusion}

We present the first systematic investigation of MIA against LALMs. While our empirical study focuses on AF3 and MF, the proposed three-phase protocol generalizes to other architectures for identifying dataset-induced distributional confounds. Our results show that the strong MIA performance is often driven by acoustic distribution shifts rather than genuine memorization. When such confounds are controlled, attack performance drops to near-random levels, suggesting strong sample-level privacy robustness in current LALMs. 

At the same time, our findings indicate that memorization may still emerge through cross-modal associations between a speaker's vocal characteristics and linguistic content. Future work should investigate which acoustic attributes, such as timbre and prosody, contribute to this phenomenon. For closed-source LALMs, the framework can be extended via shadow-model approaches~\cite{mattern2023membershipinferenceattackslanguage}, though memorization profiles may differ across model scales and modality-alignment designs. Finally, we advocate for rigorous privacy auditing that explicitly controls for distributional artifacts, paving the way toward more trustworthy and privacy-preserving audio language models.

% \section{Acknowledgments}

% {\color{blue}Acknowledgments should be included only in the camera-ready version, not in the version submitted for review. For regular papers, pages 5 and 6, and for long papers, pages 9 and 10, are reserved exclusively for acknowledgments, disclosures of the use of generative AI tools, and references. No other content may appear on these pages. Any appendices must be contained within the first four pages for regular papers and within the first eight pages for long papers.

% Acknowledgments and references may begin on an earlier page if space permits.}

% \ifcameraready
%      The Interspeech 2026 organizers
% \else
%      The authors
% \fi
% would like to thank ISCA and the organizing committees of past Interspeech conferences for their help and for kindly providing the previous version of this template.

% {\color{blue}
\section{Generative AI Use Disclosure}
During the preparation of this work, the authors utilized several generative AI tools in both the manuscript writing and experimental implementation phases. For the manuscript, ChatGPT and Google Gemini were used for assistance with language editing, rephrasing, and improving the clarity and conciseness of the text. For software development, GitHub Copilot and Cursor were employed to assist with code completion, boilerplate generation, and debugging. 

In all cases, these tools served in an assistive capacity. The core ideas, experimental design, analyses, and conclusions presented in this paper are the original work of the human authors. All authors have reviewed the final manuscript and the associated code, and take full responsibility for the integrity and correctness of the entire work.

\section{Acknowledgements}
We gratefully acknowledge the National Center for High-Performance Computing (NCHC) of Taiwan for providing the computational resources that supported this work. 
This work was also supported by the Ministry of Education (MOE) of Taiwan under the Taiwan Centers of Excellence in Artificial Intelligence program, through the NTU Artificial Intelligence Center of Research Excellence.

\bibliographystyle{IEEEtran}
\bibliography{mybib}

\end{document}